\documentclass[final,5p,12pt,twocolumn]{elsarticle}

\usepackage{epsfig}
\usepackage{lineno,hyperref}

\usepackage{threeparttable}

\biboptions{numbers,sort&compress}
\usepackage{graphics}
\usepackage{chngcntr}
\counterwithout{table}{section}

\modulolinenumbers[5]

\usepackage{amssymb}

\journal{Journal of Crystal Growth}

\biboptions{comma,square}

\begin{document}

\begin{frontmatter}



\title{Characterization and control of ZnGeN$_2$ cation lattice ordering}


\author{Eric W. Blanton}
\author{Keliang He}
\author{Jie Shan}
\author{Kathleen Kash}

\address{Department of Physics, Case Western Reserve University, 
Cleveland, OH 44106}

\begin{abstract}
ZnGeN$_2$ and other heterovalent ternary semiconductors have important potential applications in optoelectronics, but 
ordering of the cation sublattice, which can affect the band gap, lattice parameters, and phonons, is not yet well understood.  Here the effects of growth and processing conditions on the ordering of the ZnGeN$_2$ cation sublattice were investigated
using x-ray diffraction and Raman spectroscopy.  
Polycrystalline ZnGeN$_2$ was grown by exposing solid Ge to Zn and NH$_3$ vapors at temperatures between 
758 $^\circ$C and 914 $^\circ$C.  Crystallites tended to be rod-shaped, with growth rates higher along the c-axis. 
The degree of ordering, from disordered, wurtzite-like x-ray diffraction spectra to orthorhombic, 
with space group $Pna$2$_1$, increased with increasing growth temperature, as evidenced by the appearance 
of superstructure peaks and peak splittings in the diffraction patterns.  Annealing disordered, low-temperature-grown 
ZnGeN$_2$ at 850 $^\circ$C resulted in increased cation ordering.  Growth of ZnGeN$_2$ on a liquid Sn-Ge-Zn alloy at 
758 $^\circ$C showed an increase in the tendency for cation ordering at a lower growth temperature, and resulted
 in hexagonal platelet-shaped crystals. The trends shown here may help to guide understanding of the synthesis and characterization of other heterovalent ternary nitride semiconductors as well as ZnGeN$_2$.
\end{abstract}


\end{frontmatter}


\section{Introduction}

For heterovalent ternary semiconductors the cation or anion sublattice of a wurtzite or zincblende binary semiconductor is 
replaced by an ordered sublattice of two different atom types such that the average number of valence electrons stays the same.  These semiconductors often have properties similar to those of their parent binary compounds. For example, in ZnGeN$_2$ the Ga sublattice of GaN is replaced with equal numbers of Zn and Ge atoms, arranged so that in the lowest enthalpy state each N atom is bound to two Zn and two Ge atoms, and hence the average number of valence electrons remains the same; ZnGeN$_2$ is predicted to have a band gap within 100-200 meV, and lattice spacings within approximately 1-2 percent, of those of GaN \cite{Quayle15}. However, the heterovalent ternaries also have interesting and potentially useful differences in comparison to the binaries.  Because of the reduced symmetry, heterovalent ternary semiconductors can have high nonlinear optical coefficients \cite{Lambrechtbook}.  Their more complicated lattices offer more opportunities for doping strategies, and defect and band structure engineering \cite{Lambrechtbook}.  Furthermore, the degree of ordering on the cation sublattice can be used as a tuning parameter since it can change the optical properties, vibrational properties, and lattice parameters.  For example, one experimental report found that the band gap of ZnSnP$_2$ changed by 0.3 eV when the ordering parameter was changed \cite{st2010band}. In that study chalcopyrite ordering was confirmed by the appearance of sphalerite-disallowed superstructure reflections.  The greater flexibility of choice in materials provides additional opportunities. For example, ZnSnN$_2$, composed entirely of abundant elements, might replace InN or Ga$_1$$_-$$_x$In$_x$N$_2$ in some applications.  

\subsection{Cation Sublattice Ordering}

The nature of disorder in the heterovalent ternary semiconductors is of both fundamental and practical interest. However, it is not yet well understood.  One  model for disorder assumes the random placement of atoms on the mixed cation sublattice. In this model there are many instances where local charge neutrality is disrupted; the so-called octet rule is thus violated \cite{buerger1934ra1}.  These defects disrupt the electronic structure of the material, sometimes substantially reducing the band gap.  Recent theoretical work on ZnSnN$_2$ predicts that for random placement of Zn and Sn atoms on the cation sublattice, the band gap can be lower than for the ordered $Pna$2$_1$ by more than 1 eV \cite{veal2015band}. The band gap disappears altogether for the space group $Pm31$, for which the cation sublattice consists entirely of octet rule violations--each nitrogen atom is bound to either three Zn atoms and one Sn atom, or vice versa \cite{Quayle15}.

\begin{figure}
\epsfig{file=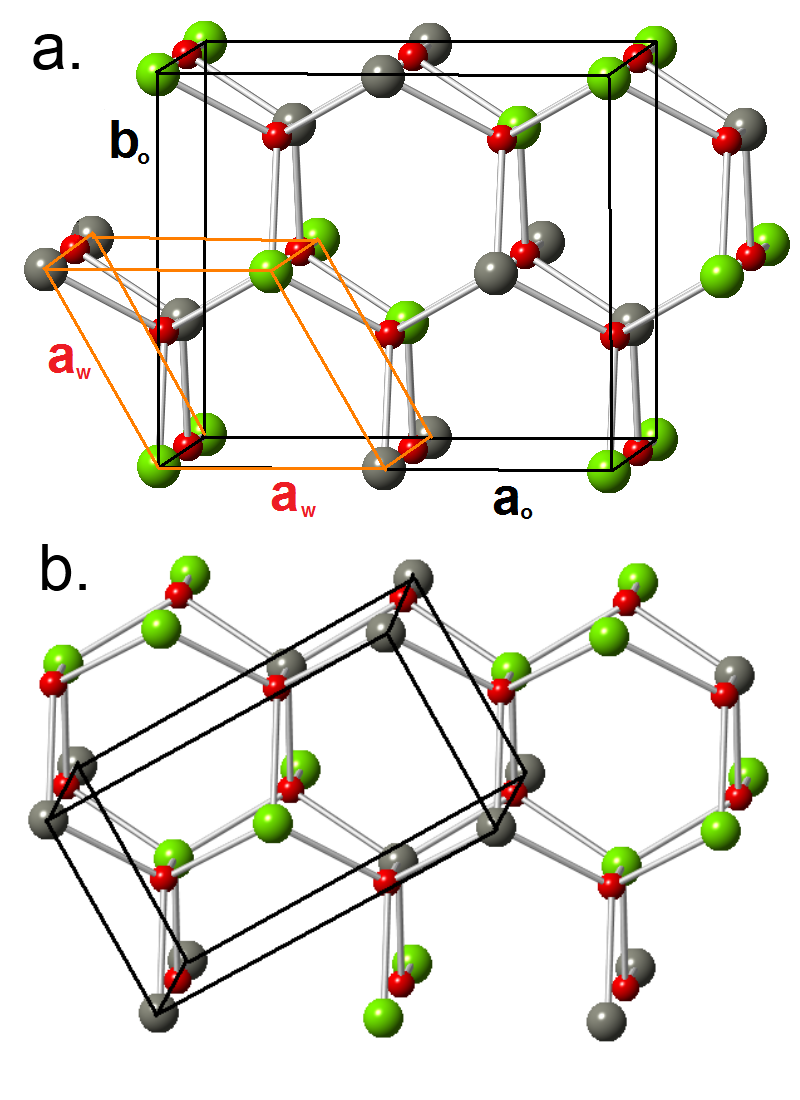,width=9cm}
\caption{(color online) a.) The unit cell of the orthorhombic $Pna$2$_1$ structure and its relation to the wurtzite unit cell.  Red spheres are nitrogen, green spheres are germanium, and grey spheres are zinc. b.) the alternate ordering arrangement structure with space group $Pmc$2$_1$. \label{unitcell} }
\end{figure}

There is another model for disorder on the cation sublattice that does not produce octet rule violations. In ZnGeN$_2$ and other wurtzite-based heterovalent ternary semiconductors, there are two ways of ordering the cation lattice that preserve the octet rule.  One configuration has the space group $Pna$2$_1$, shown Fig. \ref{unitcell}a.  The unit cell has sixteen atoms instead of four as for wurtzite, and the symmetry is orthorhombic \cite{Punya11}.  The other way of ordering has the space group $Pmc$2$_1$ and an eight-atom unit cell, \cite{lahourcade2013}, depicted in Fig. \ref{unitcell}b.  It has been shown that by randomly stacking $Pna$2$_1$- and $Pmc$2$_1$-like layers in the crystal's $y$ direction, an effectively disordered structure is formed without producing any octet-rule violations \cite{Quayle15}.  This alternative model of cation disorder could explain why ZnSnN$_2$ grown by Quayle $et$ $al.$ is observed to be disordered, as evidenced by XRD, but also has a band gap, measured by photoluminescence excitation spectroscopy and by photoluminescence spectroscopy, that is consistent with that predicted for perfectly ordered ZnSnN$_2$ \cite{ quayle2013synthesis,Quayle15}.  Similarly, there are two ways of ordering zincblende-based heterovalent ternary compounds, and a similar model of disorder has been presented for these materials \cite{wei1992first,wei1999band}.

\subsection{Detection of Ordering}

The x-ray diffraction (XRD) and Raman spectroscopy measurements used in the present study can differentiate between ordered and disordered material, but not whether the disorder obeys or violates the octet rule. 

Ordering of heterovalent ternary semiconductors is usually detected through XRD.  For ternary semiconductors derived from both zincblende and wurtzite lattices, there is often a lattice distortion associated with the cation lattice ordering.  In chalcopyrite structures $\frac{c}{a}$ can deviate from the zincblende value of 2 \cite{Garbato}.  In wurtzite-derived orthorhombic structures with space group $Pna$2$_1$, the $a$ parameter increases and the $b$ parameter decreases, causing  $\frac{a}{b}$ to become greater than the ideal wurtzite value of $\frac{2\sqrt{3}}{3}$.  These distortions, if large enough, can be detected using XRD through the splitting of many of the peaks.

In the parent binary zincblende and wurtzite structures there are reflections which are extinguished due to the symmetry of the structures.  In disordered heterovalent ternary compounds these reflections are similarly extinguished through destructive
interference arising from the disordered cation sublattice.   Superstructure peaks in the XRD pattern appear when the symmetry is reduced due to cation ordering.  The superstructure peaks can be very weak, however, because their presence depends on the difference in scattering cross section between the two cation types.  In chalcopyrite, the (101), (217), and (611) peaks are typical superstructure reflections that appear upon ordering \cite{francoeur}.  In orthorhombic $Pna$2$_1$, the (110) and (101) peaks are the strongest superstructure peaks.  In ZnGeN$_2$ these superstructure peaks are approximately two orders of magnitude less intense than the strongest peaks, while for ZnSnN$_2$, they are only one order of magnitude less intense \cite{Quayle15}.

Order-disorder transitions have been observed in many zincblende-based heterovalent ternary semiconductors.  At temperatures much lower than the melting temperatures of the compounds, the structure is chalcopyrite.  As the temperature is raised, a disorder transition is observed for some compounds, often 50-80$^\circ$C below the melting point, as evidenced by XRD.  For other compounds, the material melts before the order-disorder transition is observed \cite{Garbato}.

There has been much less research reported on ordering in the wurtzite-based heterovalent ternary compounds.  ZnGeN$_2$ and ZnSiN$_2$ \cite{Endo} have been grown with ordered $Pna$2$_1$ lattices.  ZnSnN$_2$ was first synthesized just a few years ago \cite{lahourcade2013structural,quayle2013synthesis} 
and has been receiving more interest lately as a potential earth-abundant solar absorber.  Unambiguous
observations of $Pna$2$_1$ ordering in ZnSnN$_2$ have not yet been reported.  Identification of ordering through the XRD peak-splitting is more difficult for ZnSnN$_2$ than for ZnGeN$_2$ and ZnSiN$_2$ because the amount of structure distortion due to ordering, for example the deviation of the b/a ratio from that of ideal wurtzite, is predicted to be smaller.  For ZnGeN$_2$ the b/a ratio is 2.2\% smaller than the wurtzite value, while the b/a ratio of ZnSnN$_2$ is predicted to be within 0.1\% of the wurtzite value \cite{Quayle15}. On the other hand the (110) and (101) superstructure peaks associated with $Pna$2$_1$ ordering in ZnSnN$_2$ are predicted to be an order of magnitude larger than in ZnGeN$_2$ \cite{Quayle15}, but they have not yet been reported to be observed.
  Kawamura et al. report on a high pressure synthesis method that enabled the growth of polycrystalline ZnSnN$_2$ at a higher temperature (800$^\circ$C) than has been possible before \cite{kawamura2016synthesis}.  The XRD pattern showed sharp peaks indicating high crystalline quality, but the superstructure peaks were still absent.  A recent report of ZnSnN$_2$ grown on LiGaO$_2$ showed some structural distortion that the authors attributed to 
ordering \cite{senabulya2016stab}.  Other reports show evidence of control of parameters such as the band gap and carrier concentration 
\cite{fioretti2015combinatorial,fioretti2015effects,veal2015band,feldberg2013growth} by varying growth conditions and 
by annealing. While these studies have made progress in the improvement of the quality of the ZnSnN$_2$, definitive XRD evidence of 
cation ordering is still absent.  More research is needed to determine the effects of ordering in ZnSnN$_2$ and how to control it.

A range of ordering has been observed for ZnGeN$_2$, but until recently it has not been understood how to control the cation lattice ordering \cite{Maunaye,Wintenberger,Larson74,Du,Endo,Misaki04,Peshek2008,Zhu,Kikkawa,Viennois}.  The present study shows that ZnGeN$_2$ grown at lower temperatures tends to have a high degree of disorder on the cation sublattice, while at higher temperatures the cation sublattice becomes ordered.  In addition, annealing of disordered material at high temperature causes the cation sublattice to order.  Recently, Shang et al. reported temperature-dependent ordering in ZnGeN$_2$ produced by annealing Zn$_2$GeO$_4$ under NH$_3$. Their observations are consistent with the present study \cite{Shang}.

In section 2 of this paper, the growth methods and the experiments probing the control of ZnGeN$_2$ cation lattice ordering are described.  In section 3, the XRD and Raman results demonstrating the variation and control of ordering are reported and their implications are discussed.

\section{Materials and Methods}

In this fundamental study, ZnGeN$_2$ was grown by a vapor-liquid-solid method involving the use of either a liquid Zn-Ge alloy or a liquid Zn-Ge-Sn alloy, with Sn serving in the latter case as a diluent, reacting with gaseous NH$_3$ in the center of a single-zone quartz tube furnace. While the material produced in these studies was polycrystalline, these types of growth methods in general can serve as possible routes to producing large single crystals and are therefore are of interest as potential methods for producing high quality substrates. 

In the first set of growth experiments, Zn vapor and NH$_3$ (Airgas Research Grade) flowed over a [111]-oriented Ge wafer resting on a graphite platform.  In a slightly modified second set of experiments designed to explore the effect of the addition of Sn, a Sn/Ge liquid in equilibrium with a solid Ge wafer was exposed to Zn vapor and NH$_3$. In GaN growth, the addition of diluent metals such as Na to Ga has been shown to alter the kinetics of the growth \cite{yamane1998morphology}.   For all of the experiments Zn vapor was supplied by evaporating liquid Zn (Alfa Aesar 99.999\% purity) from a graphite crucible on the upstream side of the furnace.  The Zn partial pressure was controlled by adjusting the crucible position, and thereby the crucible temperature, via a bellows-sealed plunger.  The Ge wafer platform and the Zn crucible temperatures were monitored using type-K thermocouples embedded in the graphite pieces.  The Ge wafer platform temperature was calibrated at the melting temperatures of Sn (232$^\circ$C) and Al (660$^\circ$C) by observing the change in slope of the temperature-versus-time profile associated with the heat of solidification during cooling. The temperature was calibrated to within approximately 5$^\circ$C.  H$_2$ from a hydrogen generator (Matheson Chrysalis II) was used as the carrier gas. Gas purifiers (Entegris Gatekeeper) were used at the points of injection for the H$_2$ as well as the high purity N$_2$ and NH$_3$.

\subsection{Growths on Ge wafers}
Table \ref{conditions} lists the conditions for a series of four growth experiments on Ge substrates.  Prior to being loaded into the chamber, the Ge wafer and Zn were chemically etched to reduce surface oxides.  Zn shot (Alfa Aesar 99.999\% purity) was melted into a slug, allowed to cool, then was etched in 6:1 H$_2$O:HNO$_3$ for one minute.  The Ge wafer was etched in 10\% HF for one minute.   To reduce atmospheric O$_2$ and H$_2$O contamination in the growth chamber, a series of pump-and-purge operations was performed before each experiment.  After the Zn and Ge were loaded into the quartz tube, 500 sccm of N$_2$ was flowed at 0.013 atm for 1 hour, then, the H$_2$ and NH$_3$ lines were purged by flowing 120 sccm and 60 sccm of H$_2$ and NH$_3$, respectively, at 0.013 atm for 0.5 hours.  To ensure effective purging of the chamber, including stagnant volumes, the pressure was increased to atmospheric pressure with N$_2$ then evacuated to 0.013 atm.  Five of these pump-purge cycles were performed.  Subsequently, the furnace was heated to 400$^\circ$C while flowing 50 sccm H$_2$ for approximately 16 hours.  This step further reduced atmospheric contamination and accelerated water desorption from chamber walls and reduction of oxides on the Zn and Ge surfaces.

The furnace was then raised to the growth temperature under 60 sccm of H$_2$ flow at 0.94 atm with the Zn crucible pulled out of the furnace.  The temperature was maintained for 0.5 hours to let the temperature gradients at the ends of the furnace equilibrate.  The Zn crucible was then pushed into the furnace to the location at which the desired Zn temperature was reached.  Once the Zn temperature equilibrated, the NH$_3$ flow was started in order to initiate ZnGeN$_2$ growth. For all of the experiments, the total flow rate during the growth stage was 60 sccm.  The gas flow rates were maintained during the cooling stage.

\subsection{Annealing Experiments}

For the annealing experiments, disordered ZnGeN$_2$ was grown using the conditions shown in Table \ref{conditions} for growth experiment 1.  After XRD and Raman measurements were performed, the disordered material was loaded back into the chamber for annealing.  The pump-purge cycles described in section 2.1 were performed before flowing 100 sccm N$_2$ at 0.013 atm for approximately 16 hours at room temperature, rather than at the 400$^\circ$C temperature used for the growth experiments, in order to avoid decomposition of the ZnGeN$_2$.  

At high temperature the rate of decomposition of the ZnGeN$_2$ was found to be sensitive to the ambient gas composition.  From separate experiments it was determined that if ZnGeN$_2$ was heated to 850$^\circ$C in 0.94 atm of pure N$_2$, all of the ZnGeN$_2$ would decompose within one hour, leaving pure Ge on the wafer.  Annealing at 850$^\circ$C in 0.94 atm of pure NH$_3$ resulted in complete conversion to Ge$_3$N$_4$ in one hour, as evidenced by XRD analysis.  Therefore, in order to avoid decomposition of ZnGeN$_2$ during annealing, Zn vapor and NH$_3$ were flowed while at high temperature, at the partial pressures shown in Table \ref{conditions}.  Analysis by optical microscopy before and after annealing under these conditions confirmed that the ZnGeN$_2$ crystals were unchanged in size and shape and that therefore additional ZnGeN$_2$ was not grown during the annealing experiments. As anticipated, the layer of ZnGeN$_2$ already present prevented Ge from reaching the surface to react with the ambient Zn and NH$_3$. Table \ref{conditions} lists the conditions for the two annealing experiments.

\subsection{Growth on a Ge/Sn Liquid}

A modified version of the growth method described in section 2.1 consisted of growing ZnGeN$_2$ on a Ge/Sn liquid.  A shaving of Sn (Alfa Aesar 99.999\%) approximately 0.05 g in mass was placed on the Ge wafer prior to being loaded into the furnace.  Upon heating to the growth temperature of 758$^\circ$C, a liquid alloy formed that was in equilibrium with the solid Ge wafer.  After the furnace reached the desired temperature, the Zn crucible was pushed into the furnace and heated to the desired temperature.  As the Zn partial pressure was increased, Zn dissolved into the Ge/Sn liquid.  The Ge/Sn liquid was left exposed to the Zn pressure for 15 minutes in order for the liquid composition to equilibrate before flowing NH$_3$.  At the growth conditions, the liquid composition was estimated to be approximately 31 at.\% Sn, 15 at.\% Zn, and 54 at.\% Ge, based on thermodynamic data \cite{lee,long,olesinski}. 

\subsection{Characterization Methods}

Powder x-ray diffraction measurements were done with a Bruker Discover D8 with a 2-dimensional detector.

For Raman spectroscopy measurements, either a 633 nm or a 532 nm wavelength laser beam was sent through a 50x objective onto the sample.  The beam was defocused so that an area approximately 70 $\mu$m wide was illuminated.  The scattered light was collected in reflection by the same objective, filtered, and detected by a spectrometer equipped with 
a liquid-nitrogen-cooled CCD.

\section{Results}

\subsection{Growths on Ge wafers}
 
\begin{figure*}[t]
\includegraphics[width=18cm]{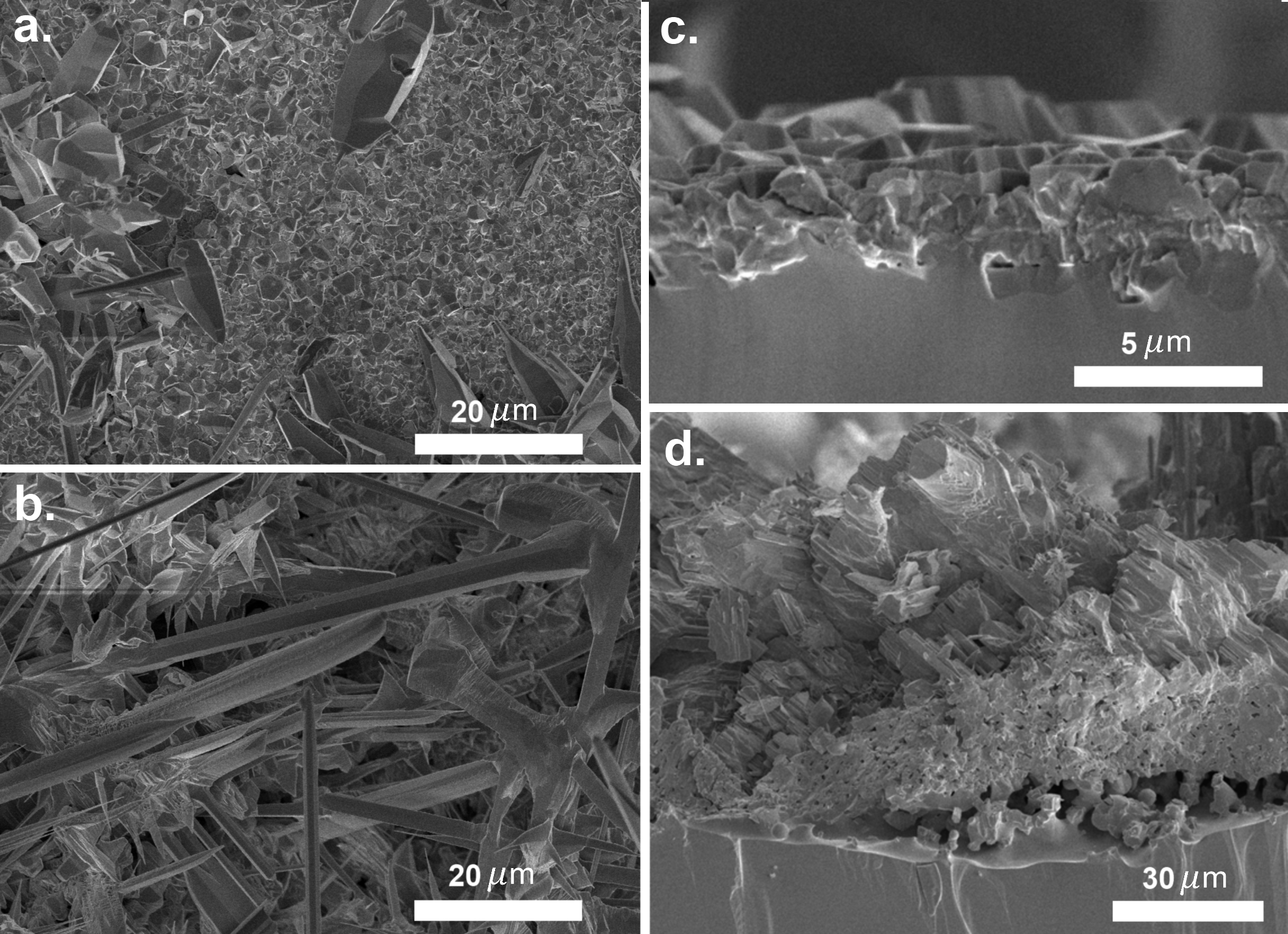}
\caption{SEM micrographs of the growths on Ge wafers. a.) surface of growth experiment 1, grown at 758$^\circ$C. b.) surface of growth experiment 3, 
grown at 852 $^\circ$C. c.) cross section, after cleaving, of growth experiment 1. d.) cross section, after cleaving, of growth experiment 4, grown at
 914 $^\circ$C \label{sem} }
\end{figure*}

\begin{figure*}
\epsfig{file=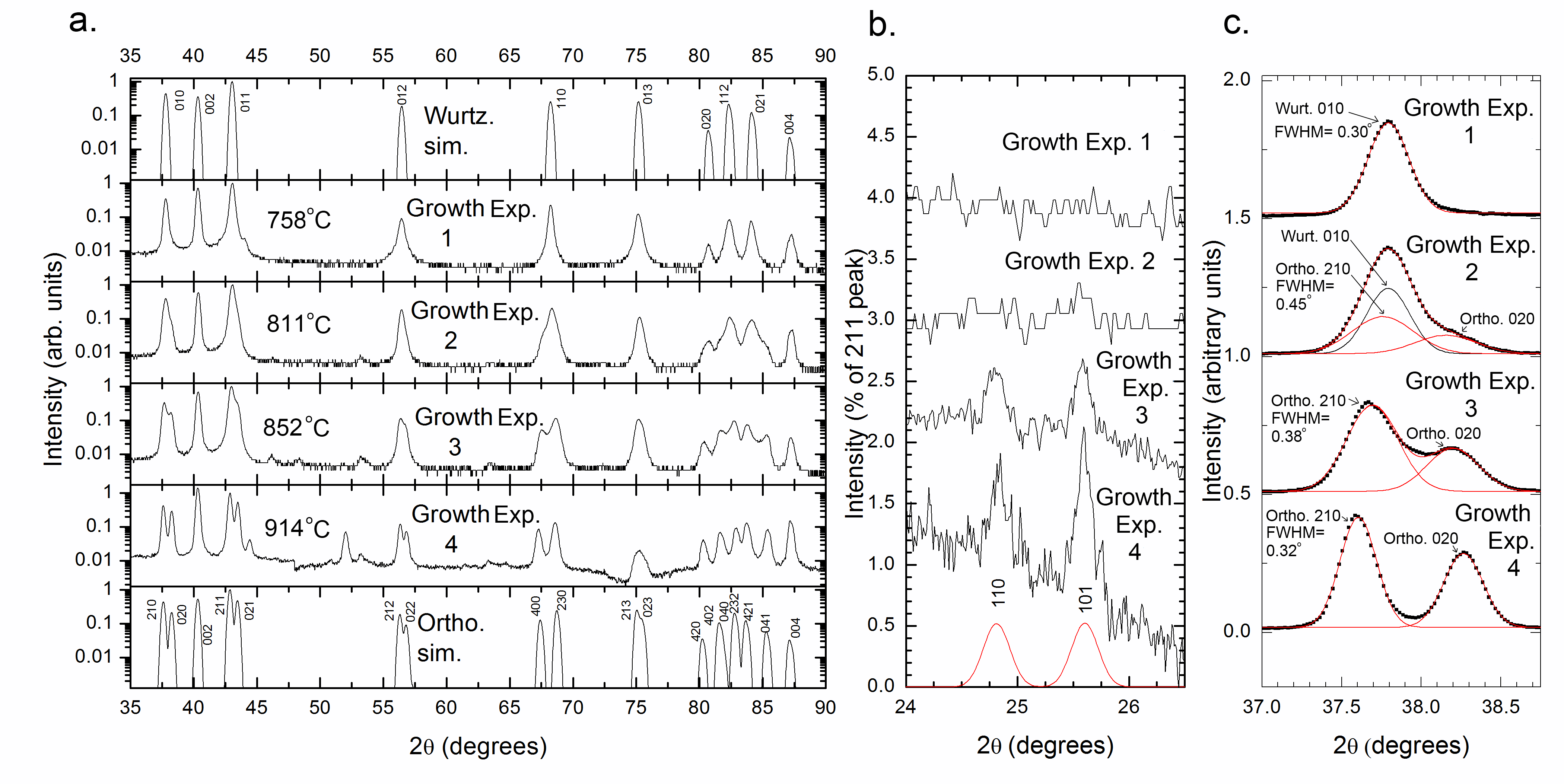,width=18cm}
\caption{XRD patterns of the four growths on Ge wafers. Co K$\alpha$1 x-rays were used.  a.) The pattern at the top is the simulated wurtzite pattern with each cation lattice site occupied by 50\% Ge and 50\% Zn \cite{crystaldiffract}.  The pattern at the bottom is the simulated orthorhombic pattern for the space group $Pna$2$_1$ using the lattice parameters found for growth experiment 4 in Table \ref{parameters}.
 b.) The portions of the patterns showing the (110) and (101) superstructure peaks.  c.) 
The portions of the patterns showing 
the (210) and (020) peaks.  The fitted FWHM of the (210) peak is listed for each growth experiment. \label{waferxrd} }
\end{figure*}

Figures \ref{sem}$a$ and \ref{sem}$b$ show SEM micrographs of the surfaces of growth experiments 1 and 3, respectively. Figures \ref{sem}$c$ and \ref{sem}$d$ show the cross-sections of the cleaved wafers of growth experiments 1 and 4, respectively.  The smooth area at the bottom of the frame in figures \ref{sem}$c$ and \ref{sem}$d$  is the underlying Ge wafer.  The crust thickness tended to become thicker and the crystallite sizes tended to be larger as the growth temperature was increased.

Fig. \ref{waferxrd} shows the XRD patterns for the four growth experiments.  The simulated patterns were made using Crystal Diffract software using lattice parameters fit from the measured patterns, listed in Table \ref{parameters}.  The wurtzite simulation was calculated assuming each cation site had an average occupancy of 50\% Ge and 50\% Zn atoms.  The XRD pattern of growth experiment 1, at the lowest growth temperature, is indistinguishable from that of wurtzite, consistent with this material having a mostly disordered cation sublattice.  As the growth temperature was increased, the cation ordering increased, as evidenced by the increased peak splitting observed in the figure.  Fig. \ref{waferxrd}c shows the (210) and (020) peaks with an expanded x-axis.  The patterns were fit using Gaussian curves in order to derive the lattice parameters.  It is clear that the peak splitting increased with growth temperature, but the resulting pattern was not always fit well by a single phase.  Since there are twice the number of (210) planes as (020) planes in ordered ZnGeN$_2$, the (210) peak should be twice the height of the (020) peak.  Growth experiment 2 does not have this 2:1 ratio.  Instead, the pattern of growth experiment 2 was fit well by a mixture of an ordered orthorhombic phase and a disordered wurtzite phase.  In the fit, the orthorhombic phase was constrained to have a 2:1 intensity ratio for the (210) and (020) peaks, and the wurtzite phase was constrained to have the same lattice parameter and peak width as the disordered material of growth experiment 1.  Researchers studying ZnSnP$_2$ also found evidence of phase separation of the ordered chalcopyrite and disordered sphalerite structures \cite{st2010band}.  Growth experiments 3 and 4 were each fit well by the single orthorhombic $Pna$2$_1$ phase.

\begin{table*}
{\scriptsize
\caption{Growth temperatures, gas partial pressures, and durations of the experiments.  For all experiments, the total pressure was maintained at 0.94 atm and the total flow rate was maintained at 60 sccm.}\label{conditions}}
\begin{tabular}{ccccccc}
\hline
Expt. & Growth Temp. ($^0$C) & P$_Z$$_n$ (atm) & P$_N$$_H$$_3$ (atm) & P$_H$$_2$ (atm) & P$_N$$_2$ (atm) & Duration (hours)   \\ \hline
Growth exp. 1   & 758 & 0.027 & 0.31 & 0.63 & 0 & 3.7  \\
Growth exp. 2   & 811 & 0.056 & 0.24 & 0.70 & 0 & 4.0  \\
Growth exp. 3   & 852 & 0.028 & 0.31 & 0.63 & 0 & 1.5  \\
Growth exp. 4   & 914 & 0.028 & 0.47 & 0.47 & 0 & 2.0 \\
Anneal 1   & 850 & 0.03 & 0.47 & 0 & 0.47 & 1.0  \\

Anneal 2   & 850 & 0.03 & 0.47 & 0 & 0.47 & 4.0  \\

Platelet   & 758 & 0.03 & 0.31 & 0.63 & 0 & 4.0   \\

   	\hline
\end{tabular}
{\footnotesize
}

\end{table*}

Equilibrium thermodynamics predicts that the ordered phase is stable at low temperature and the disordered phase is stable at high temperature.  The trend observed here with ZnGeN$_2$ is opposite to that predicted by thermodynamics, so we conclude that the equilibrium disorder temperature is higher than the temperatures explored here, and the observed disordered samples are in metastable states.  We speculate that at the lower growth temperatures atoms do not have sufficient kinetic energy to overcome barriers to finding the lowest-energy, ordered configuration, while at higher temperatures they do.

The orthorhombic $a$ and $b$ lattice parameters change as the growth temperature is increased, as indicated by the shifting peaks in Fig. \ref{waferxrd}.  The fitted lattice parameters are shown in Table \ref{parameters}.  As the growth temperature increases from that of growth experiment 1, 758$^\circ$C, increasing fractions of the cation sublattice become ordered.  In the beginning stages of ordering we should expect to see some mixing of ordered and disordered material, which is what we see with growth experiment 2.  In these mixtures, regions of ordered ZnGeN$_2$ are in close proximity to regions of disordered ZnGeN$_2$, which has different $a$ and $b$ parameters, so the $a$ and $b$ lattice parameters may not be fully relaxed.  As the proportion of ordered material increases, the $a$ and $b$ parameters may be able to more fully relax, resulting in the $a$ parameter growing and the $b$ parameter shrinking with increasing growth temperature.  The inhomogeneous strain that may result from the mixing of ordered and disordered material may be the cause here of the broader XRD peaks observed for growth experiments 2 and 3.  As shown in Fig. \ref{waferxrd}c, the deconvolved  $Pna$2$_1$ (210) peak is absent for growth experiment 1, widest for growth experiment 2, then narrows as the material becomes more completely ordered at higher growth temperatures.  Modeling of powder diffraction patterns in CrystalDiffract was done to estimate the strain in the samples.  For growth experiment 1, a strain of 0.11 (+/-0.03) \% was consistent with the change in peak width with scattering angle.  For growth experiment 3, a strain of 0.18 (+/-0.05) \% was found.  For the highest temperature growth experiment 4, a strain of 0.11 (+/-0.03) \% was found.

\begin{table}
{\scriptsize
\caption{The fitted lattice parameters.  The parameter $a$ was calculated using the 
(210) and (020) peaks, $b$ using the (020) peak, and $c$ using the (002) peak. The numbers in parentheses are the calculated uncertainties in the last place digits.}\label{parameters}}
\begin{tabular}{p{1.7cm} p{1.4cm} p{1.4cm} p{1.4cm} }
\hline
Expt. & a (\AA) & b (\AA) & c (\AA) \\ \hline
Growth exp. 1   & 6.383$(5)$ & 5.523$(4)$ & 5.192$(4)$ \\
Growth exp. 2   & 6.41$(1)$ & 5.477$(6)$ & 5.189$(4)$ \\
Growth exp. 3   & 6.43$(1)$ & 5.468$(9)$ & 5.190$(4)$ \\
Growth exp. 4   & 6.450$(4)$ & 5.462$(4)$ & 5.193$(4)$ \\
Anneal 1   & 6.42$(2)$ & 5.486$(7)$ & 5.190$(4)$ \\

Anneal 2   & 6.43$(1)$ & 5.476$(6)$ & 5.190$(4)$ \\

Platelet   & 6.44$(1)$ & 5.468$(6)$ & 5.191$(4)$  \\

   	\hline
\end{tabular}

\end{table}

\begin{figure}
\includegraphics[width=9cm]{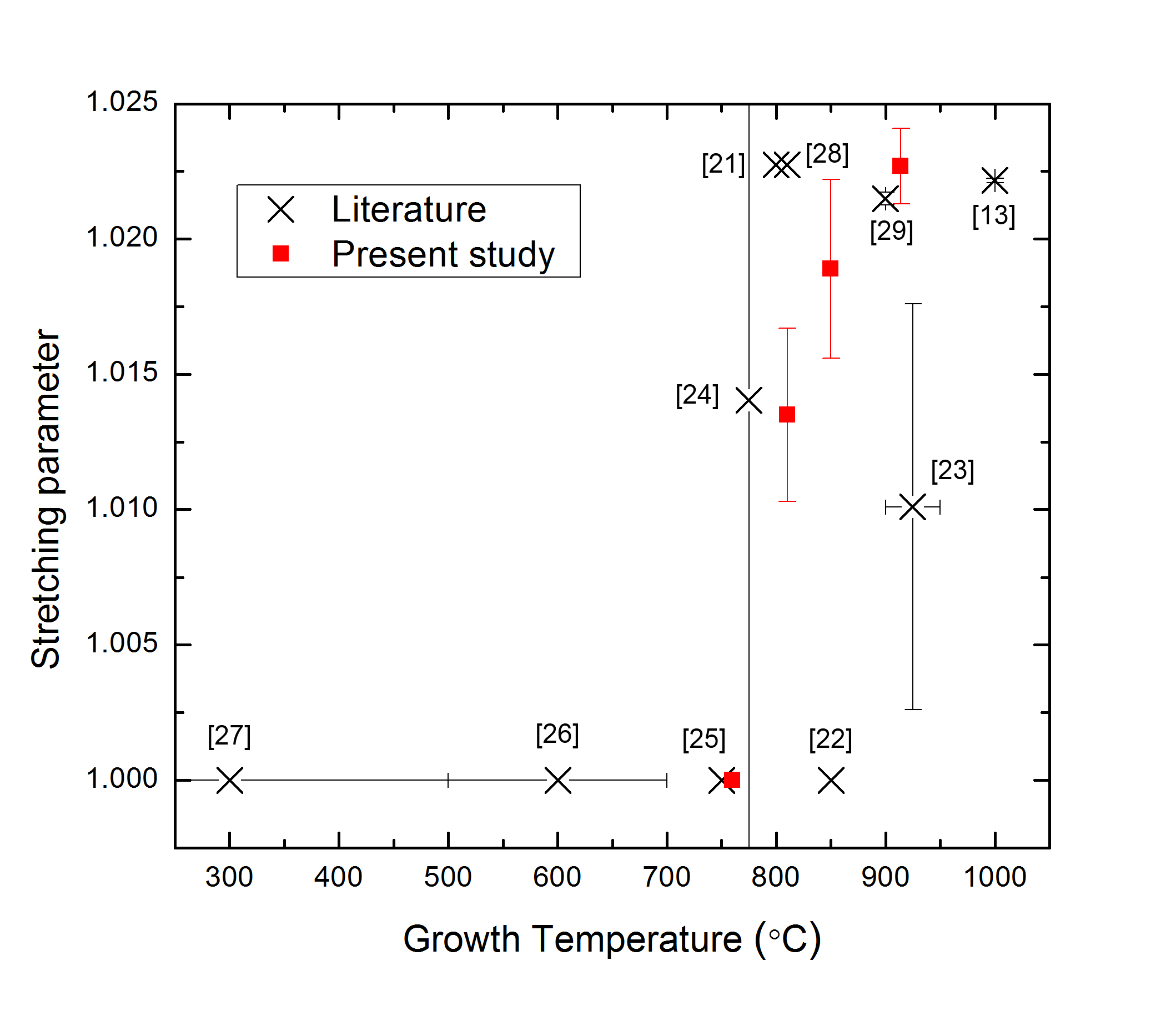} 
\caption{The relation between cation lattice ordering and growth temperature for the ZnGeN$_2$ reported here and for ten reported growths in the literature.  The stretching parameter $\alpha = \frac{3a}{2b\sqrt{3}}$ is a measure of the orthorhombic lattice distortion and is used here as a measure of the extent of ordering.  The references from which the data were taken are shown in brackets. \label{stretch}}
\end{figure}

Here we compare the lattice parameters of ZnGeN$_2$ in the literature to our results, as functions of growth temperature.  We introduce the stretching parameter, $\alpha=$$\frac{3a}{2b\sqrt{3}}$, where $a$ and $b$ are the orthorhombic lattice parameters, 
 to quantify the orthorhombic cation lattice ordering. A wurtzite structure viewed in the orthorhombic unit cell has an $\frac{a}{b}$ ratio of  $\frac{2\sqrt{3}}{3}$, so disordered material has $\alpha$ equal to 1, and partially or fully ordered material has $\alpha$ greater than 1.  Fig. \ref{stretch} plots the stretching parameter $\alpha$ for the four growths on Ge wafers in the present study and for ten ZnGeN$_2$ growths reported in the literature, as a function of growth temperature.  All of the ZnGeN$_2$ lattice parameters reported in the literature were converted to the orthorhombic system for ease of comparison.  Below growth temperatures of approximately 760$^\circ$C, all reported material is disordered (wurtzite-like).  Above this temperature, almost all material is reported to have some amount of orthorhombic ordering.

Fig. \ref{waferxrd}b shows, for the material reported here, the portion of the XRD pattern containing the (110) and (101) superstructure peaks.  The pattern for growth experiment 1 shows no superstructure peaks, consistent with that material being mostly disordered.  The pattern of growth experiment 2 shows small peaks consistent with partial ordering.  The patterns of growth experiments 3 and 4 show (110) and (101) peak heights consistent with the prediction for fully ordered $Pna$2$_1$ material, using atomic positions from \cite{Punya11}.  To our knowledge measurement of these superstructure peaks for ZnGeN$_2$ using XRD has not been reported by any other group to date.

\begin{figure}
\epsfig{file=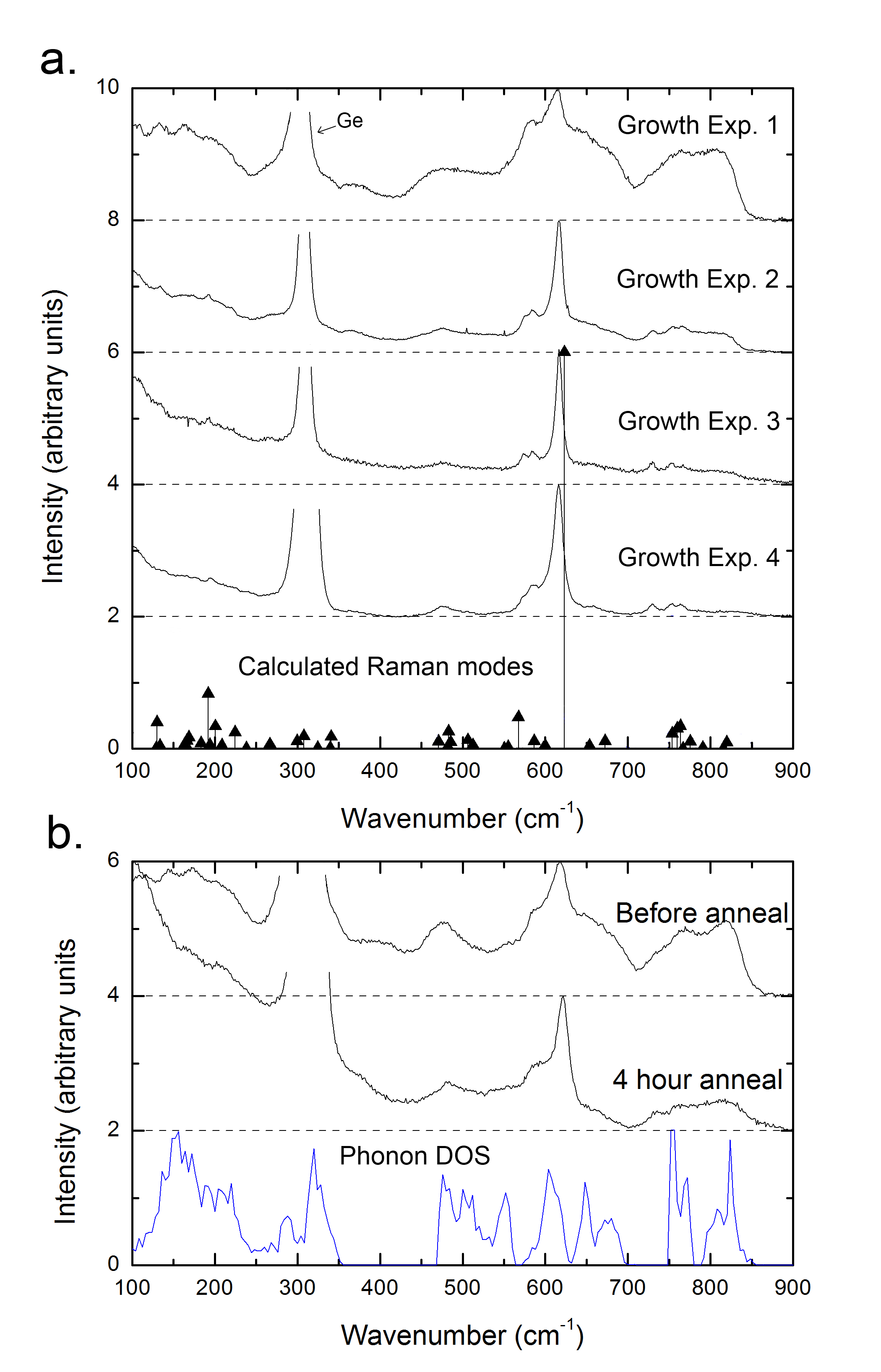,width=9cm}
\caption{(color online) Raman spectra for polycrystalline ZnGeN$_2$  a) for the four growths on Ge wafers and b) after annealing the disordered material for four hours.  The triangles at the bottom of a) are the calculated Raman-active modes of ZnGeN$_2$ \cite{Paudel08}.  In b), the blue
 curve at the bottom of the graph is the calculated phonon density of states \cite{Paudel08}. \label{raman}}
\end{figure}

Disorder on the cation lattice disrupts the phonon modes of 
ordered ZnGeN$_2$, and therefore Raman spectroscopy can be used as an additional method to probe the disorder.  In a previous study on single crystal ZnGeN$_2$, strong DOS features were observed in the Raman spectra and were interpreted as arising from the partial breakdown of momentum conservation rules due to disorder \cite{Peshek08}.  The trends observed here are consistent with this interpretation.
 The micro-Raman spectra for the four growth experiments are shown in Fig. \ref{raman}.  The spectra are normalized so that the large peak at 616 cm$^{-1}$ in each spectrum has the same intensity relative to the background above 850 cm$^{-1}$.  As the growth temperature is decreased, the sharpness of the observed peaks decreases and the phonon DOS features increase in intensity.  As the disorder increases, the momentum conservation rules are relaxed and phonons from across the Brillouin zone participate in the scattering.  The trend toward greater ordering with higher growth temperature observed here in the Raman spectra is consistent with that observed in the XRD patterns.

\subsection{Annealing Experiments}

\begin{figure}
\includegraphics[width=9cm]{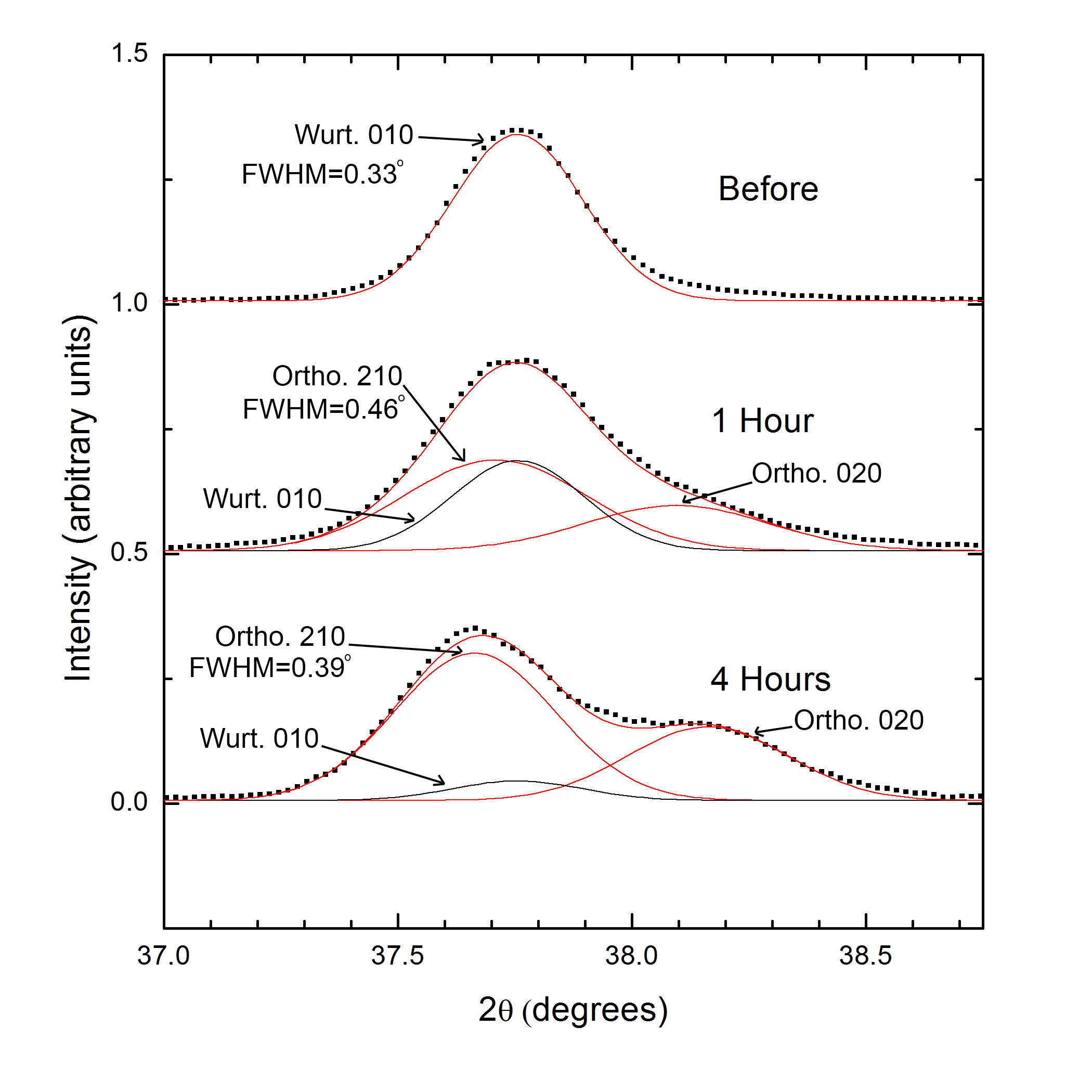} 
\caption{Portion of the XRD pattern showing the (210) and (020) peaks for disordered as-grown ZnGeN$_2$ and after one or four hours of annealing at 850$^\circ$C.\label{annealing}}
\end{figure}

Disordered ZnGeN$_2$ samples were annealed at 850$^\circ$C for either one or four hours in the presence of Zn and NH$_3$ vapor.  Fig. \ref{annealing} shows the (210) and (020) XRD peaks for the material before and after the annealing experiments.  The first pattern is fit well by wurtzite, consistent with the material being mostly disordered.  Peak splitting is evident after one hour of annealing and increases after four hours, consistent with the material becoming increasingly ordered with longer annealing time.  The peaks were fit with Gaussian curves to find the lattice parameters as for the analysis described in section 3.1.  

Fig. \ref{raman}b shows the micro-Raman spectra for the disordered material and after four hours of annealing.  The trend is similar to that of the four growths on Ge wafers in Fig. \ref{raman}a.

\subsection{Platelets}

\begin{figure}
\includegraphics[width=9cm]{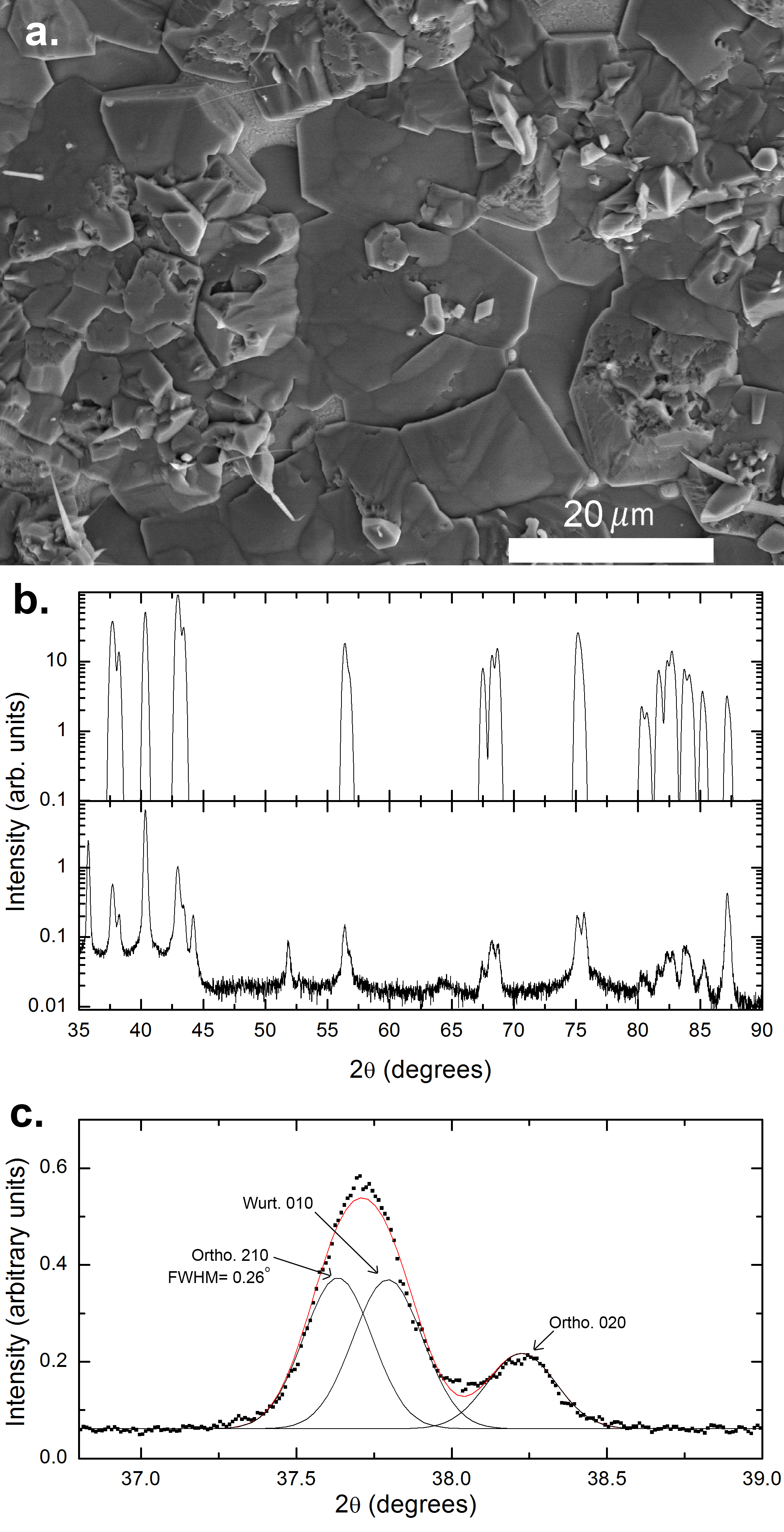} 
\caption{a.) SEM micrograph of ZnGeN$_2$ platelets grown on a Ge-Sn-Zn liquid.  b.) measured XRD pattern of the platelets and a simulated pattern of a mixture of 66\% ordered orthorhombic ZnGeN$_2$ and 34\% disordered wurtzite ZnGeN$_2$. c.) portion of the XRD pattern showing the (210) and (020) peaks.  The (210) peak width is shown. \label{platelet}}
\end{figure}

Both the ordering and morphology of ZnGeN$_2$ were affected substantially when Sn was added to the Ge wafer. Without Sn, the growth rate along the $c$ axis was more than an order of magnitude larger than along the $a$ and $b$ axes, and hexagonally faceted rod-shaped structures resulted. When Sn was added the growth rate along the $a$ and $b$ axes was one to two orders of magnitude higher than along the $c$ axis, resulting in platelet-like structures.  Fig. \ref{platelet} a. shows a SEM micrograph of the ZnGeN$_2$ platelets.  The specific reasons for the change in morphology with the addition of Sn are not understood at this time.

In addition to ZnGeN$_2$, some trace amounts of ZnO grew as evidenced in the XRD pattern.  ZnO was not present in any of the growths done without Sn present.  We hypothesize that the presence of Sn catalyzes the growth of ZnO from remaining impurities.  The ZnO was removed by etching the sample in a 2.5\% HCl:water solution for six minutes.  The XRD pattern shown in Fig. \ref{platelet} was taken after the ZnO was removed.

Growing on the Sn liquid resulted in cation lattice ordering at a lower growth temperature when compared with the growths directly on Ge wafers.  Fig. \ref{platelet}b shows the XRD pattern of the platelets, grown at 758$^\circ$C, which is the same temperature as growth experiment 1 on a Ge wafer in section 3.1.  The platelets are much more ordered, as evidenced by the XRD pattern, while the material grown at the same temperature directly on a Ge wafer (Growth experiment 1) is mostly disordered.  The XRD pattern of the platelet crystals is a mixture of ordered orthorhombic and disordered wurtzite phases.  The peaks in Fig. \ref{platelet} were fit according to the procedure in section 3.1.  During the XRD measurement the sample was rotated in order to assure random sampling of the planes perpendicular to the c-plane.  The specific reasons for the change in the tendency to order are not understood at this time but might be related to the change in preferential growth direction.

ZnGeN$_2$ and ZnSnN$_2$ can in principle be alloyed together, and alloys have been synthesized by RF sputtering 
\cite{narang2014bandgap}.  Therefore, we must consider the possibility that the material grown in the presence of Sn is a ZnGeN$_2$-ZnSnN$_2$ alloy.  Comparing the $c$ parameter of the platelets in Table \ref{parameters} with the other samples reveals that they are equal within the uncertainty of the measurements.  Using Vegard’s law, the uncertainty in the $c$ parameter of $\pm$0.004 $\AA$ is equivalent to a ZnSnN$_2$ volume fraction of 0$\pm$1.5\%.

\section{Conclusions}

These synthesis experiments showed that it is possible to control the cation lattice ordering of ZnGeN$_2$ by varying the growth temperature and by annealing.  The material was disordered at the lowest temperature growth experiment (758$^\circ$C) and became progressively more ordered as the growth temperature was increased to 914$^\circ$C.  The equilibrium disorder transition temperature is most likely significantly higher than the temperatures explored here, and so the disordered samples observed here are most likely in metastable states.  Growth or annealing at higher temperatures resulted in material closer to the equilibrium ordered $Pna$2$_1$ state.  Ordering was detected using the peak splitting and appearance of superstructure peaks in the measured XRD patterns.  The trend between disorder and increase of phonon DOS features in Raman spectra was consistent with the trend observed with the XRD patterns.

Growth on a Sn-based liquid decreased the temperature at which ordered ZnGeN$_2$ was produced.  This modified growth method also changed the crystal morphology, resulting in hexagonal platelets instead of hexagonally faceted rod-shaped crystals.

\section{Acknowledgements}

E.B. and K.K. were supported by the National Science Foundation grants DMR-1006132 and DMR-1409346. J.S. and K.H. acknowledge support from the National Science Foundation grant DMR-1106225.

\section*{References}

\bibliography{zgn}

\end{document}